\documentclass{elsart3}
\usepackage{graphicx}

\begin{document}

\begin{frontmatter}

\title{Pseudogap and Competing States in Underdoped Cuprates}

\author{Patrick A.  Lee\thanksref{thank1}},

\address{Center for Materials Science and Engineering and
Department of Physics, MIT, Cambridge, MA 02139 USA}

\thanks[thank1] {E-mail: palee@mit.edu}



\begin{abstract}
I shall argue that the high $T_c$ problem is the problem of doping into a Mott insulator. Furthermore, the well
documented pseudo-gap phenomenon in underdoped cuprates holds the key to understanding this physics. Phase
fluctuation alone cannot explain this phenomenon, but  there is a clear need to identify  a competing state which
lives in the vortex core. The staggered flux state is a good candidate for the competing state and experimental
tests of these ideas will be discussed.
\end{abstract}

\end{frontmatter}

In the past several years, a concensus has begun to emerge that the phenomenon of high temperature
superconductivity in cuprates is associated with doping into a Mott insulator.  The undoped
material is an antiferromagnet with a large exchange energy $J$ of order 1500~K.  The doped holes hop with a
matrix element $t$, which is estimated to be approximately $3J$.  However, the  N\'{e}el state is not favorable
for hole hopping, because after one hop the spin finds itself in a ferromagnetic environment.  Thus it
is clear that the physics is that of competition between the exchange energy $J$ and the hole
kinetic energy per hole $xt$.  Apparently the superconducting state emerges as the best
compromise, but how and why this occurs is the central question of the high T$_c$ puzzle.  In
the underdoped region this competition results in physical properties that are most anomalous. 
The metallic state above the superconducting T$_c$ behaves in a way unlike anything we have
encountered before.   Essentially, an energy gap appears in some properties and not others, and
this metallic state has been referred to as the pseudogap state.  We will focus our attention
on this region because the phenomenology is well established and we have the best chance of
sorting out the fundamental physics.  

The pseudogap phenomenon is most clearly seen in the uniform susceptibility.  For example,
Knight shift measurement in the YBCO 124 compound shows that while the spin susceptibility $\chi_s$ is
almost temperature independent between 700~K and 300~K, as in an ordinary metal, it decreases
below 300~K and by the time the T$_c$ of 80~K is reached, the system has lost 80\% of the spin
susceptibility.\cite{1}  
To emphasize the universality of this phenomenon, I reproduce in Fig. 1 some old data on YBCO and LSCO.  Figure 1(a)
shows the Knight shft data from Alloul {\it et al}. from 1989.\cite{2}  I have subtracted the orbital
contribution, which is generally agreed to be 150~ppm,\cite{3} and drawn in the zero line to highlight the spin
contribution to the Knight shift which is proportional to $\chi_s$.  The proportionality constant is known\cite{2}
which allows us to draw in the Knight shift which corresponds to the 2D square $S = {1\over 2}$ Heisenberg
antiferromagnet with
$J = 0.13$~eV.\cite{4,5}  The point of this exercist is to show that in the underdoped region, the spin 
susceptibility drops {\it below} that of the Heisenberg model at low temperatures before the onset of
superconductivity.  This trend continues even in the severely underdoped limit (O$_{0.53}$ to O$_{0.41}$), showing
that the $\chi_s$ reduction cannot simply be understood as fluctuations towards the antiferromagnet. 
Note that the discrepancy is worse if $J$ were replaced by a smaller $J_{eff}$ due to doping, since $\chi_s \sim
J_{eff}^{-1}$. 
The data seen
in this light strongly point to singlet formation as the origin of the pseudogap seen in the uniform spin
susceptibility.

\begin{figure}[h]
\includegraphics[width=1.0125\linewidth]{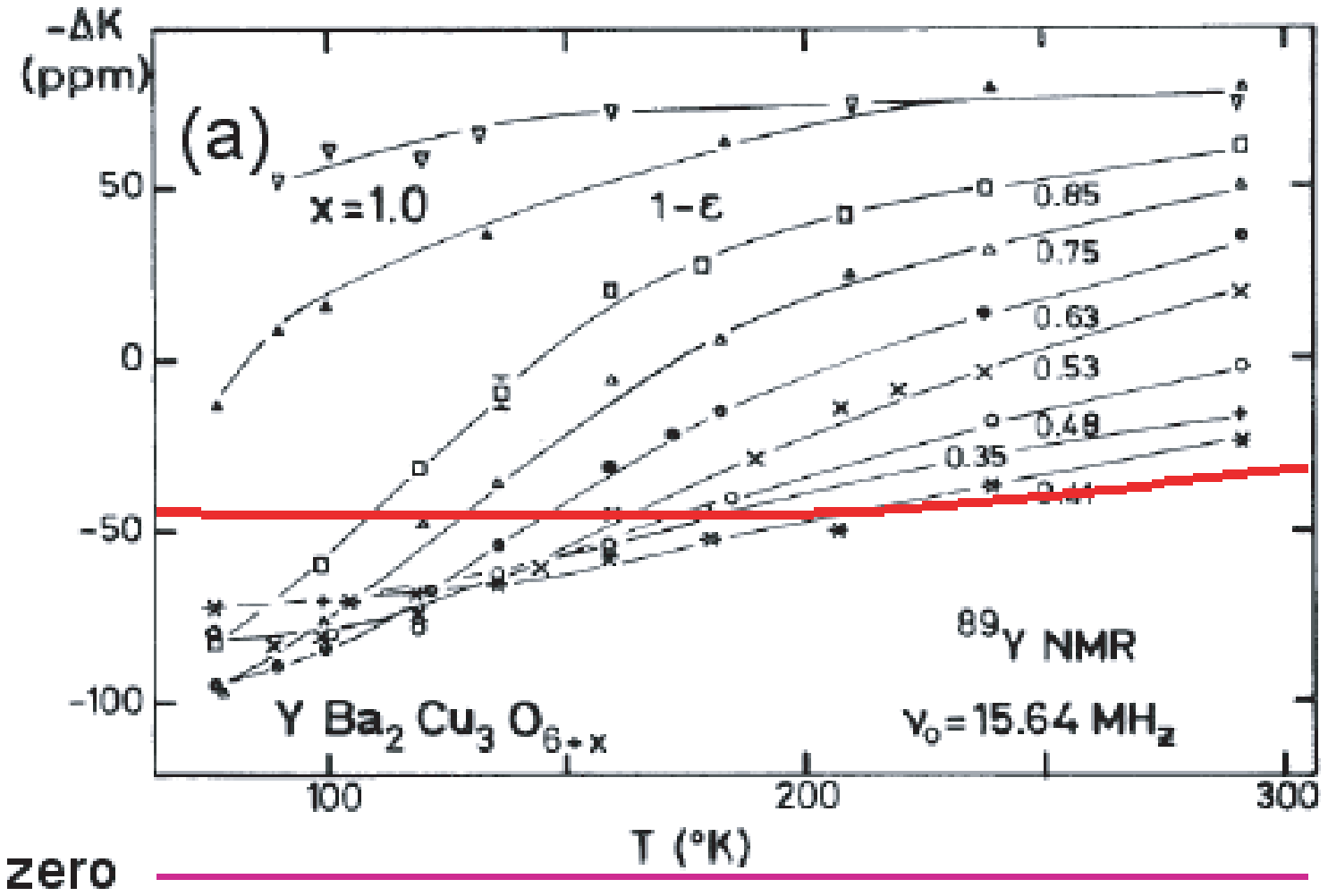}
\includegraphics[width=1.0325\linewidth]{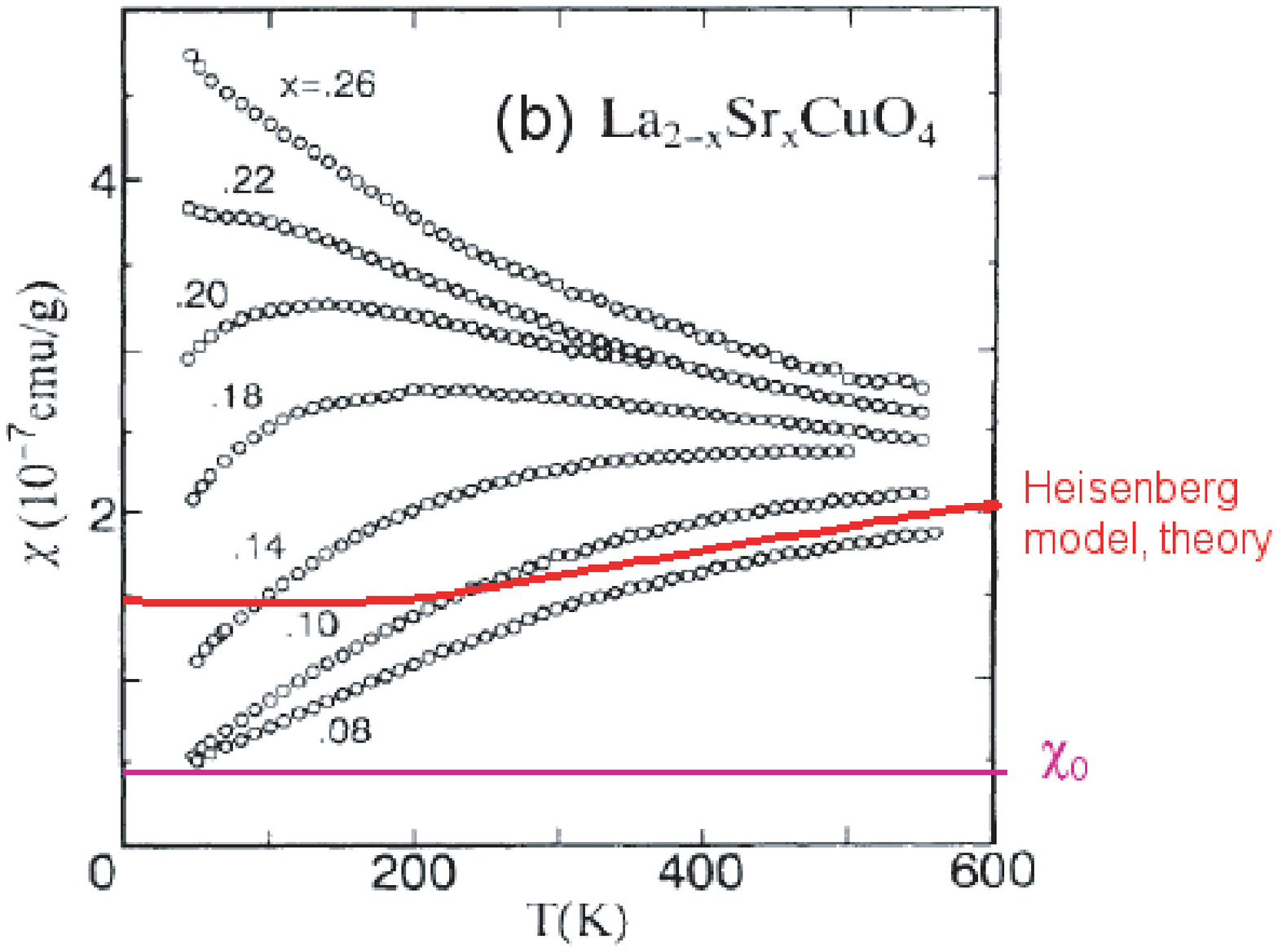}
\caption{(a) Knight shift data of YBCO for a variety of doping.\cite{2}  The zero level for the spin
contribution has been added and the solid line represents the prediction of the 2D $S={1\over 2}$
Heisenberg model for $J = 0.13$~eV.  (b) Uniform magnetic susceptibility for LSCO.\cite{7}  The orbital
contribution $\chi_0$  is shown (see text) and the solid line represents the Heisenberg model
prediction.}
\end{figure}

It is worth noting that the trend shown in Fig. 1 is not so apparent if one looks at the measured spin
susceptibility directly.\cite{6}  This is because the van~Vleck part of the spin susceptibility is doping
dependent, due to the changing chain contribution.  This problem does not arise for LSCO, and in Fig. 1(b) we show
the uniform susceptibility data.\cite{7}  The zero of the spin part is determined by comparing susceptibility
measurements to $^{17}$O Knight shift data.\cite{8}  Nakano {\it et al}.\cite{7} find an excellent fit for the $x =
0.15$ sample (see Fig. 9 of ref.  7) and determine the orbital contribution for this sample to be $\chi_0 \sim 0.4
\times 10^{-7}$~emu/g. This again allows us to plot the theoretical prediction for the Heisenberg model.\cite{9} 
Just as for YBCO, $\chi_s$ for the underdoped samples ($x = $0.1 and 0.08) drops below that of the Heisenberg model. 
In fact, the behavior of
$\chi_s$ for the two systems is remarkably similar, especially in the underdoped region.

A second indication of the pseudogap comes from the linear $T$ coefficient of the specific heat, which shows a marked
decrease below room temperature.\cite{10}  It is apparent that the spins are forming into singlets and
the spin entropy is gradually lost.  On the other hand, the frequency dependent conductivity
behaves very differently depending on whether the electric field is in the $ab$ plane
$(\sigma_{ab})$ or perpendicular to it $(\sigma_c)$.  At low frequencies (below 500~cm$^{-1}$)
$(\sigma_{ab})$ shows a typical Drude-like behavior for a metal with a width which decreases
with temperature, but an area (spectral weight) which is independent of temperature.\cite{11}  Thus
there is no sign of the pseudogap in the spectral weight.  This is surprising because in other
examples where an energy gap appears in a metal, such as the onset of charge or spin density
waves, there is a redistribution of the spectral weight from the Drude part to higher
frequencies.  On the other hand below 300~K $\sigma_c(\omega)$ is gradually reduced for frequencies below
500~cm$^{-1}$ and a deep hole is carved out of $\sigma_c(\omega)$ by the time T$_c$ is
reached.\cite{12}   Finally, angle-resolved photoemission shows that an energy gap (in
the form of a pulling back of the leading edge of the electronic spectrum from the Fermi
energy) is observed near momentum $(0,\pi)$ and the onset of superconductivity is marked by the
appearance of a small coherent peak at this gap edge.

The pseudogap phenomenology is well explained by a cartoon picture which emerges from the RVB
(resonating valence band) theory of Anderson.\cite{13}  The spins are paired into singlet pairs. 
However, the pairs are  not static but are fluctuating due to quantum mechanical superposition,
hence the term quantum spin liquid.  The singlet formation explains the appearance of the spin gap and the
reduction of spin entropy.  The doped holes appear as vacancies in the background of singlet
pair liquid and can carry a current without any energy gap.  However in $c$-axis conductivity
and electron is removed from one plane and placed on the next.  The intermediate state is an
electron which carries spin 1/2 and therefore it is necessary to break a singlet pair and pay
the spin-gap energy.  The same consideration applies to the photoemission experiment.  Finally, according to RVB
theory, superconductivity emerges when the holes become phase coherent.  The spin singlet
familiar in the BCS theory has already been formed.

While the above picture is appealing, there has been another popular explanation.  The idea is that the pseudogap
phenomenology can be understood as a superconductor with robust amplitude but strong phase fluctuation.  The
superfluid density
$\rho_s$ which controls the phase stiffness is proportional to the doping concentration $x$ and
becomes small in the underdoped region.  As emphasized by Uemura\cite{4} and by Emery and Kivelson,\cite{5}
T$_c$ is controlled by $\rho_s$ and is much lower than the energy gap.  We shall now argue that
phase fluctuations cannot be the whole story.  Setting aside the question of where
the strong pairing amplitude comes from in the first place, that the phase fluctuation scenairo
is incomplete can be seen from the following argument.  In two dimensions the destruction of
superconducting order is via the Berezinskii-Kosterlitz-Thouless (BKT) theory of vortex
unbinding.  Above T$_c$ the number of vortices proliferate and the normal metallic state is
reached only when the vortex density is so high that the cores overlap.  At lower
vortex density, transport properties will resemble a superconductor in the flux flow regime. 
In ordinary superconductors, the BKT temperature is close to the mean field temperature, and
the core energy rapidly becomes small.  However, in the present case, it is postulated that the
mean field temperature is high, so that a large core energy is expected.  Indeed, in a
conventional core the order parameter and energy gap vanish, costing $\Delta^2_0/E_F$ per unit area of energy. 
Using a core radius of $\xi = V_F/\Delta_0$, the core energy of a conventional superconductor is $E_F$.  In our
case, we may replace $E_F$ by $J$.  If this were the case, the proliferation of vortices would not happen until a
high temperature $\sim J$ independent of $x$ is reached.  Thus for the phase fluctuation scenario to to work, it
is essential to have ``cheap'' vortices, with energy cost of order T$_c$.  Then the essential problem is to
understand what the vortex core is made of.  Put another way, there has to be a competing state with energy very
close to the $d$-wave superconductors which constitute the core.  The vortex core indeed offers a glimpse of the
normal state reached when $H$ exceeds
$H_{c2}$, and is an important constituent of the pseudogap state above T$_c$.

What are the candidates for the competing order?  A candidate which has attracted a lot of
attention is the stripe phase.\cite{16,17}    In the LSCO familty, dynamical stripes (spin
density waves) are clearly important, especially near $x = {1\over 8}$.  There are recent
report of incommensurate SDW nucleating around vortices.  However, until now there has been little
evidence for stripes outside of the LSCO family.  On the theoretical side, as a competing state
it is not clear how the stripes are connected to $d$-wave superconductivity and it is hard to
understand how the nodal quasiparticles turn out to be most sharply defined on the
Fermi surface, since these have to transverse the stripes at a $45^\circ$ angle.

A second candidate for the vortex core is the antiferromagnetic state.  This possibility was
first proposed several years ago in the context of the SO(5) theory.\cite{18}  This theory is
phenomenological in that it involves only bosonic degrees of freedom (the SDW and pairing order
parameters).  The quasiparticles are out of the picture.  Thus the fundamental question of
how the holes are accommodated has not really been addressed.   There are reports of enhanced antiferromagnetic spin
fluctuations, and perhaps even static order, using NMR.\cite{19,20}  I shall argue next that other considerations
also lead to antiferromagnetic fluctuations and possibly static orders inside the vortex core, so that the
observation of antiferromagnetic cores does not necessarily imply the existence of SO(5)
symmetry.

Finally, I come to the candidate which we favor --- the staggered flux state with orbital
currents.\cite{21}  Indeed, using the staggered flux state as the core, Lee and Wen have successfully
constructed a ``cheap'' vortex state.\cite{22}  The staggered flux phase has an
advantage over other possibilities in that its excitation spectrum is similar to the $d$-wave
superconductor and the SU(2) theory allows us to smoothly connect  it to the
superconductivity.   We also regard the staggered flux phase as the precursor to  N\'{e}el
order, so that antiferromagnetic  fluctuations or even SDW order are accommodated naturally. 
Of course, it is experiments which have the final say as to which candidate turns out to be
realized.  Our strategy is to work out as many experimental  consequences as we can and propose
experiments to confirm or falsify our theory.

The staggered flux state was first introduced as a mean field solution at half-filling\cite{23}
and later was extended to include finite doping.\cite{24}   At half-filling, due to the constraint of no double
occupation, the staggered flux state corresponds to an insulating state with power law decay in the spin
correlation function.  It is known that upon including gauge fluctuations which enforce the
constraint, the phenomenon of confinement and chiral symmetry breaking occurs, which directly
corresponds to  N\'{e}el ordering.\cite{25}  The idea is that with doping, confinement is
suppressed at some intermediate energy scale, due to screening by holes and to dissipation.\cite{26}
As the temperature is lowered,  the pseudogap state emerges which can be understood as fluctuating
between the staggered flux state and the $d$-wave superconducting state.  
As still lower tamperature, the staggered flux states become dilute and form the core of fluctuating $hc/2e$
vortices. Finally,  the vortices bind via the BKT transition and 
the $d$-wave superconducting state is the stable ground state. 
Thus the staggered flux state may be regarded as the ``mother state'' which is an unstable
fixed point due to gauge fluctuations.  It flows to  N\'{e}el ordering at half-filling and to
the $d$-wave superconductor for sufficiently large $x$.  Thus the staggered flux state plays a
central role in this kind of theory.  This picture is depicted schematically in Fig. 2.  
We should point out that the
staggered flux state (called the $D$-density wave state) has recently been proposed as the ordered state in the
pseudogap region.\cite{27}  As explained elsewhere,\cite{28} we think that this view is not supported
by experiment and we continue to favor the fluctuation picture.

\begin{figure}[h]
\begin{center}\leavevmode
\includegraphics{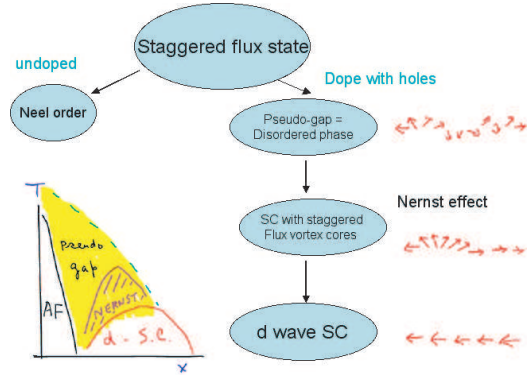}
\caption{Schematic representation of our view of the phase diagram.  In the undoped limit the staggered flux
state develops into the N\'{e}el state once gauge fluctuations are included.  The
pseudogap phase
can be thought of as a fluctuating phase between the staggered flux state and the $d$-wave
superconductor.  The
SU(2) theory allows us to smoothly connect these fluctuations.  In this theory different states may be
represented
by a three-dimensional quantization axis with arrows pointing to the north and south poles representing
staggered
flux order and arrows in the equator representing the  $d$-wave superconductor.  These arrows are
schematically
shown in the figure.  As temperature is lowered, the staggered flux regions (arrows pointing to north
or south pole)
are localized to form cores of $hc/2e$~vortices.  Eventually these vortices disappear via the
Kosterlitz-Thouless
transition to the
$d$-wave superconducting state.  There is a broad range in  temperatures above this transition where
vortices may
give rise to Nernst effects.\cite{39}  This is sketched in the schematic phase diagram.
}
\end{center}
\end{figure}

The above picture finds support from studies of  projected wavefunctions, where the 
no-double-occupation constraint is enforced by hand on a computer.  With doping the best state
is a projected
$d$-wave state.   This state can explain many of the properties of the superconductor, as recently discussed by
Paramekanti {\it et al}.\cite{29}  It is natural to consider the projected staggered flux state (at finite doping) as
a trial wavefunction for the ``normal state'' which exists inside the vortex core.  The energy difference between
this and the projected $d$-wave superconductor may be considered the condensation energy.  The condensation energy per
site computed this way is shown in Fig. 3.\cite{30}  Note the dome shape which is reminiscent of the $T_c$ curve
and the rather small value for the condensation energy, consistent  with our expectation based on the SU(2) idea. 
Another interesting quantity we calculated\cite{31} is the current-current correlation function for the projected
$d$-wave BCS wavefunction:
$
c_j(k,\ell) = < j(k)j(\ell) >
$
where $j(k)$ is the physical electron current on the bond $k$.  The average current $<j(k)>$ is
obviously zero, but the correlator exhibits a staggered circulating pattern.\cite{31} 
Such a pattern is absent in the $d$-wave BCS state before
projection, and is a result of the Gutzwiller projection.  Our result for $c_j$ is consistent with exact
diagonalization of two holes in 32 sites.\cite{32}

The staggered current generates a staggered physical
magnetic field (estimated to be 10--40 gauss)\cite{24,28} which may be detected, in principle, by neutron
scattering.  In practice the small signal makes this a difficult, though not impossible experiment and we
are motivated to look for situations where the orbital current may become static or quasi-static.  Recently, we
analyzed the structure of the $hc/2e$ vortex in the superconducting state within the SU(2) theory and concluded
that in the vicinity of the vortex core, the orbital current becomes quasi-static, with a time scale determined by
the tunnelling between two degenerate staggered flux states.\cite{22}  It is very likely that this time is
long on the neutron time scale.   Thus we propose that a quasi-static peak centered around $(\pi,\pi)$ will appear
in neutron scattering in a magnetic field, with intensity proportional to the number of vortices.  The time
scale may actually be long enough for the small magnetic fields  generated by the orbital currents to be
detectable by
$\mu$-SR or Yttrium NMR.  Again, the signal should be proportional to the external fields.  (The NMR experiment
must be carried out in 2--4--7 or 3 layer samples to avoid the cancellation between bi-layers.)  We have also
computed the tunnelling density of states in the vicinity of the vortex core, and predicted a rather specific kind
of period doubling which should be detectable by atomic resolution STM.\cite{33}  The recent
report\cite{34} of a static field of $\pm 18$ gauss in underdoped YBCO which appears in the vortex state is
promising, even though muon cannot distinguish between orbital current or spin as the origin of the magnetic
field.  We remark that in the underdoped antiferromagnet, the local moment gives rise to a field of 340 gauss
at the muon site.  Thus if the 18 gauss signal is due to spin, it will correspond to roughly $1/20$th of the
full moment.

\begin{figure}[h]
\begin{center}\leavevmode
\includegraphics[width=0.8\linewidth]{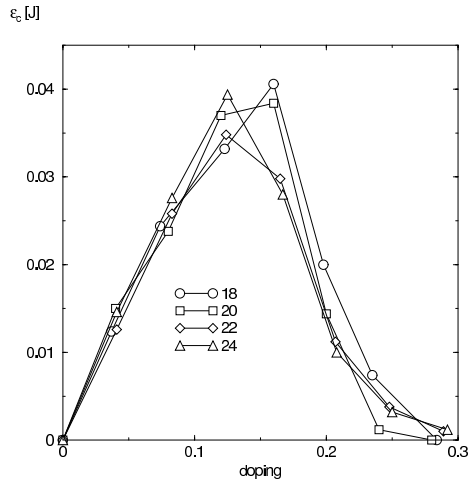}
\caption{The condensation energy per site vs. doping
as estimated from the difference between the energy of the projected staggered flux state and the
projected $d$-wave superconductor.\cite{30}  Data are shown for a variety of
sample sizes.}
\end{center}
\end{figure}

We remark that our analytic model of the vortex core is in full agreement with the numerical solution of
unrestricted mean field $\Delta_{ij}$ and $\chi_{ij}$ by Wang, Han and Lee\cite{35} and Ogata and
collaborators.\cite{36} Recently we found that for small doping in the $t$-$J$ model a small moment SDW co-exists
with orbital currents in the vortex core.\cite{37}  More generally, we expect $(\pi,\pi)$ spin fluctuations to be
enhanced\cite{38} so that the tendency to antiferromagnetism is fully compatible with the staggered flux picture. 
This vortex solution is also interesting in that the tunnelling density of states show a gap, with no sign of the
large resonance associated with Caroli-deGennes-type core levels found in the standard BCS model of the vortex. 
 The low density of states inside the vortex core has an important implication.  In the standard
Bardeen-Stephen model of flux-flow resistivity, the friction coefficient of a moving vortex is due to dissipation
associated with the vortex core states.  Now that the core states are absent, we can expect anomalously small
friction coefficients for underdoped cuprates.  The vortex moves fast transverse to the current and gives rise to
large flux-flow resistivity.   Since the total conductivity is the sum of the flux-flow conductivity and the
quasiparticle conductivity, it is possible to get into a situation where the quasiparticle conductivity dominates
even for H
$\ll$ H$_{c2}$.   Thus the ``cheap'' and ``fast'' vortex opens the possibility of having vortex states above the
nominal T$_c$ and H$_{c2}$, when the resistivity looks like that of a metal, with little sign of flux-flow
contribution.  From this point of view, the large Nerst effect observed by Ong and
co-workers \cite{39} over a large region in the H-T plane above the nominal T$_c$ and H$_{c2}$ (as determined
by resistivity) may be qualitatively explained.  The schematic phase diagram is shown in Fig. 2.

I thank X.-G. Wen, N.  Nagaosa, D. Ivanov, J. Kishine and Y. Morita for their collaboration on the
work reviewed here.  I acknowledge the support of NSF through the MRSEC program with grant number DMR 02--13282.

\end{document}